\documentclass[aps,prl,showpacs,twocolumn]{revtex4}
\usepackage{amsmath,amssymb,amsthm,amscd,amsfonts}
\usepackage{graphicx}
\usepackage{colordvi}
\usepackage{color}

\newcommand{\PT}{{$\cal PT$}}
\newcommand{\si}{\sigma_i}
\newcommand{\sr}{\sigma_r}

\begin{document}

\title{Solitons in \PT-symmetric nonlinear lattices }

\author{Fatkhulla Kh. Abdullaev$^{1}$,  Yaroslav V.  Kartashov$^2$,   Vladimir V. Konotop$^{1}$, and Dmitry A. Zezyulin$^{1}$}

\affiliation{ $^1$Centro de F\'isica Te\'orica e Computacional, Faculdade de Ci\^encias,
Universidade de Lisboa,   Avenida Professor
Gama Pinto 2, Lisboa 1649-003, Portugal
\\
 $^2$ICFO-Institut de Ciencies Fotoniques, and Universitat Politecnica de Catalunya, Mediterranean Technology Park, 08860 Castelldefels (Barcelona), Spain
}

\date{\today}

\begin{abstract}
Existence of localized modes supported by the \PT-symmetric nonlinear lattices is reported. The system considered reveals  unusual properties: unlike other typical dissipative systems it possesses families (branches) of solutions, which can be parametrized by the propagation constant;  relatively narrow localized modes appear to be stable, even when the conservative nonlinear lattice potential is absent; finally, the system supports stable multipole solutions.
\end{abstract}

\maketitle


Since introduction of the concept of the \PT-symmetric potentials~\cite{first}, this subject attracted a great deal of attention~\cite{special}. While the primary interest was devoted to such systems in the context of non-Hermitian quantum mechanics, recently new  applications of the \PT-symmetric potentials have been found in optics in media with inhomogeneous in space gain and damping, i.e. with properly designed imaginary part of the linear refractive index. The first experiments reporting the phenomenon are already available~\cite{experim}. As soon as the importance of the optical applications was realized, it became also clear that the phenomenon can be studied in the nonlinear context,
from the point of view of existence of nonlinear localized modes in linear \PT-symmetric potentials~\cite{Christo1}.
It is then natural step to address the existence and stability of localized modes in {\em nonlinear} \PT-symmetric potential which in optics can be implemented by means of proper spatial modulation of   nonlinear gain and losses. As an example, such optical systems can be nonlinear waveguides, employing concantenated  semiconductor optical amplifier and semiconductor doped two-photonic absorber sections (notice that experimental implementation of the linear \PT symmetry breaking was reported in~\cite{experiment}).

While it is now known that stable localized~\cite{Malomed,Sivan} and moving~\cite{AG}  solitons can exist in conservative purely nonlinear lattices~\cite{Malomed,Sivan,AG}  (see also \cite{CBKS,KMT} for review)  the existence of stable localized solitons in complex nonlinear lattices is still an open problem, since up to now only periodic waves were found to be stable in such structures~\cite{AKSY}. The elucidation of stable localized solitons in \PT-symmetric nonlinear lattices is therefore a central goal of this Letter.

We describe the propagation of laser radiation along the $\xi$-axis of the medium with periodic transverse modulation of cubic nonlinearity and nonlinear gain with the complex  nonlinear Schr\"odinger (NLS) equation for the dimensionless light field amplitude $q$:
\begin{eqnarray}
\label{CNLS}
i q_\xi =-\frac 12 q_{\eta\eta}- |q|^2 q-\left[V(\eta)+ i W(\eta)\right]|q|^2q
\end{eqnarray}
where $\eta$ and $\xi$ are the normalized transverse and longitudinal coordinates, respectively. The functions $V(\eta)$ and $W(\eta)$ describe transverse periodic modulations of the conservative and dissipative parts of the nonlinearity and are assumed to satisfy the \PT symmetry relations.
We further assume that conservative and dissipative parts of nonlinearity have the same period $\pi$, i.e.   $V(\eta)=V(-\eta)= V(\eta+\pi)$ and $W(\eta)=-W(-\eta)=W(\eta+\pi)$.  These functions will  be considered bounded with $\sigma_r$ and $\si$ being the maxima of $V$ and $W$, respectively.

We are interested in stationary localized solutions, which can be searched in the form $q(\eta,\xi)=u(\eta)e^{i\theta(\eta)+ib\xi}=[w_r(\eta)+iw_i(\eta)]e^{ib\xi}$ where $u$ and $\theta$ are the amplitude and phase of the mode. Eq.~(\ref{CNLS}) can be rewritten in the hydrodynamic form
\begin{equation}
\label{hydro}
\frac 12 u_{\eta\eta}-b u+[1+V(\eta)]u^3- \frac{j^2}{2u^3}=0,\,\,\, j_\eta=- 2W(\eta)u^4
\end{equation}
where we have introduced the "current density" given by $j=u^2d\theta/d\eta$.

Starting with general properties of  stationary localized solutions, we notice that it follows from (\ref{hydro}) that such solutions can exist only for $b>0$ and their asymptotical behavior at $\eta\to\pm\infty$ is given by $u\sim e^{\pm\sqrt{2b}\eta}$ and   $j\sim e^{\pm 4\sqrt{2b}\eta}$, i.e. the current density is localized much stronger than the field.
One easily finds that the field $u$ can become zero only in the points where the current density $j$ is zero, as well. Furthermore, closely following the approach described in~\cite{CBKS}, one obtains that $b={\cal O}(U^2)$ where  $U=U_r+U_i$, with $U_{r,i}=\int_{-\infty}^{\infty}w_{r,i}^2(\eta)d\eta$, is the total energy flow of the beam. This implies that in the limit of small intensity ($U\to0$) we have $U\sim\sqrt{b}$, and respectively $ u_{max}\to 0 $. Following~\cite{CBKS}, one can obtain  the relation $b\leq u_{max}^2(1+\si)$. Thus the existence of the  solutions with $u_{max}\to 0 $ implies $b\to 0$.

\begin{figure}[htb]
\vspace{-4.2cm}
\includegraphics[width=\columnwidth]{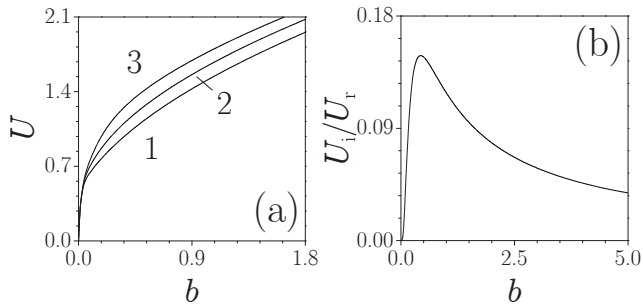}
\vspace{-4.2cm}
\caption{ (Color online)  (a) Energy flow versus propagation constant for fundamental solitons at $\sigma_i=0$ (curve 1), $1.4$ (curve 2), $1.8$ (curve 3) for $\sigma_r=1$. (b) $U_i/U_r$ versus propagation constant at $\sigma_i=1.8$, $\sigma_r=1$.}
\label{fig2}
\end{figure}

The energy balance in stationary solutions follows from (\ref{hydro}) and reads $\int_{-\infty}^{\infty}W(\eta)u^4(\eta)d\eta=0$. Since we consider odd functions $W(\eta)$, this condition can be satisfied by any even function $u(\eta)$. In other words, unlike this happens in dissipative systems of a general kind~\cite{Akhmediev}, the requirement of balance between losses and gain in our system does not introduce a constraint selecting only one possible mode (i.e. the propagation constant is not determined by the balance between losses and gain). Thus, in terms of the existence of branches of  solutions the properties of Eq.~(\ref{CNLS})  resemble the properties of the conservative NLS where the propagation constant $b$ is determined by the energy flow $U$ and a continuous family of solutions exist. This fact is illustrated  in Fig.~\ref{fig2}. Hereafter in all numerical simulations we use 
\begin{eqnarray}
\label{VW}
V(\eta)=\sr\cos(2\eta)\quad\mbox{and}\quad W(\eta)=-\si\sin(2\eta),
\end{eqnarray}
 where $\sr$ and $\si$ are the modulation depths of the conservative and dissipative lattices.
In Fig.~\ref{fig2}~(a) we observe that increase of $b$  results in monotonic growth of the soliton  energy flow $U$
and the contraction of light in a single channel of nonlinear lattice. Now, however the energy of the soliton is distributed between real and imaginary components of the field, as it is shown in Fig.~\ref{fig2}~(b). The ratio $U_i/U_r$ of energy flows concentrated in imaginary and real parts of the field takes on maximal value at intermediate $b$  values and diminishes at  $b\to 0$  and $b\to \infty$.

Let us now turn to more detailed study of the mentioned limits of the propagation constant. First of all, Fig.~\ref{fig2}~(a) supports the above estimate $U\sim b^{1/2}$, also illustrating that in  the limit $b\to0$ the energy flow very weakly
depends on the amplitude of dissipative part of potential (the three lines are indistinguishable on the scale of the picture).  Indeed, in this case
the lattice period becomes small in comparison with the width of smoothly modulated soliton
 and one can perform the averaging procedure~\cite{Malomed}. For the model (\ref{VW}) the solution of (\ref{CNLS}) can be found in the form
 $
q(\eta,\xi) \approx Q(\eta,\xi) + A(\eta,\xi)\cos(2\eta) + iB(\eta,\xi)\sin(2\eta) ,
$
where $Q$, $A$, and $B$ are the  functions slowly varying on the  scale $\pi$.
Substitution of this ansatz in Eq.~(\ref{CNLS}) yields $A= \sr |Q|^2 Q/2$ and  $B  = -\si|Q|^2 Q$, and cubic-quintic NLS equation for the field $Q$:
\begin{eqnarray}
\label{CQNLS}
iQ_\xi +\frac{1}{2}Q_{\eta\eta} +  |Q|^2Q +  \frac{3}{2}\chi |Q|^4 Q =0.
\end{eqnarray}
where $\chi=\frac{1}{6}(3\sr^2 - \si^2)$
 This equation does not contain any imaginary part - the consequence of the opposite parities of real and imaginary components of the nonlinearity modulations. The solitonic solution of (\ref{CQNLS}) which exists at $b\chi>-1/8$ is well known (see e.g. \cite{Pushk}).
\begin{equation}
\label{QCNLS_solit}
Q=  2\sqrt{b}e^{ib \xi}\left[1 + \sqrt{1 + 8\chi b}\cosh(2\sqrt{2b}\eta)\right]^{-1/2}.
\end{equation}
This solution is reduced to the conventional NLS soliton in the limit $b\to0$, revealing weak dependence of the soliton on the parameter $\chi$, what explains convergence of all branches in Fig.~\ref{fig2}~(a) at $b\to 0$.

\begin{figure}[htb]
\vspace{-1.5cm}
\includegraphics[width=\columnwidth]{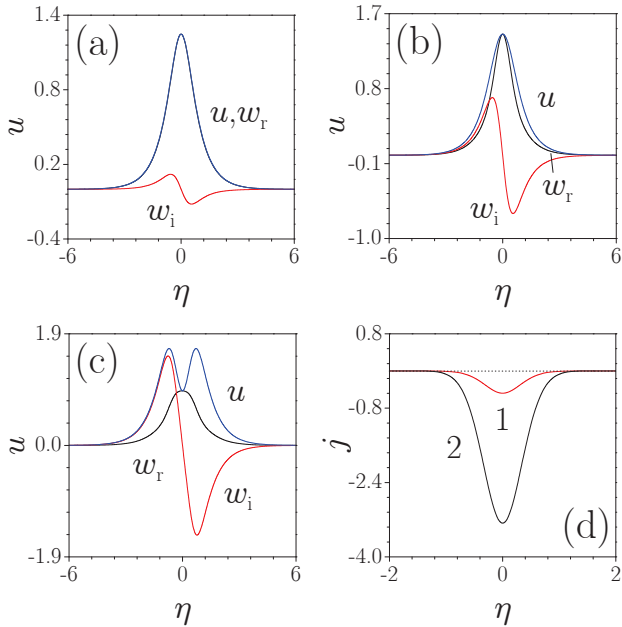}
\vspace{-2cm}
\caption{ (Color online)  The profiles of fundamental solitons from lower branch at $\sigma_i=0.5$ (a) and $1.58$ (b). (c) The profile of soliton from higher branch at $\sigma_i=0.3$. (d) The current density for the fundamental solitons from lower branch at $\sigma_i=0.5$ (curve 1) and $1.58$ (curve 2). In all cases $b=1$ and $\sigma_r= 0.5$}
\label{fig1}
\end{figure}

\begin{figure}[htb]
\vspace{-2.2cm}
\includegraphics[width=0.9\columnwidth]{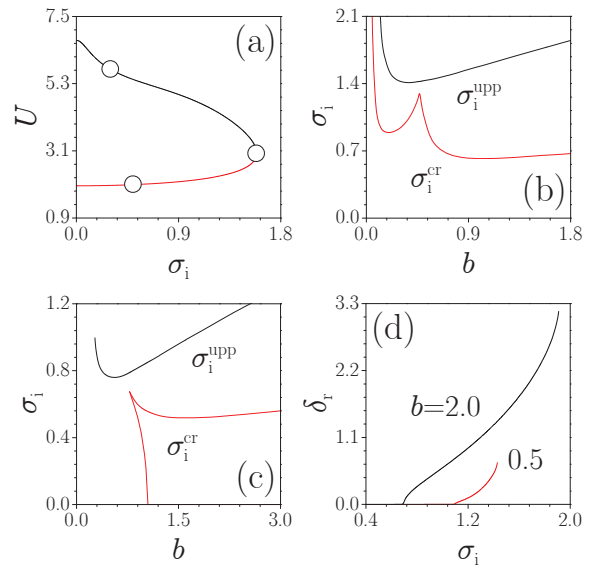}
\vspace{-1.5cm}
\caption{ (Color online)  (a) The energy flow {\it vs} $\si$ for lower (red curve) and upper (black curve) branches of fundamental solitons at $b=1$, $\sr=0.5$. Circles correspond to solitons shown in Figs.~\ref{fig1}(a)--(c). Domains of existence and stability on the plane $(b,\si)$  for fundamental solitons at $\sr=0.5$ (b) and  $\sr=0$ (c). (d) The perturbation growth rate {\it vs} $\si$ at $\sr=0.5$.}
\label{fig3}
\end{figure}

The profiles of the simplest fundamental soliton solutions of Eq.~(\ref{CNLS}), i.e. the solitons belonging to the lowest branch (see also Fig.~\ref{fig3} (a), below), are shown in Fig.~\ref{fig1} (the phase of the solution  is fixed by the condition  $\theta(0)=0$, it however can be changed due to the phase invariance of the complex NLS equation). The centers of such solitons reside in the point where conservative part of nonlinearity takes on the maximal value, while dissipative part of nonlinearity is zero. Due to the fact that left wing of soliton resides in the domain with nonlinear losses, while its right wing is subjected to nonlinear gain the solitons are characterized by the anti-symmetric imaginary parts of the field [Figs.~\ref{fig1}(a) and  (b)] indicating on tilted phase fronts and the existence of internal currents directed into the domain with losses [Fig.~\ref{fig1} (d)].

Returning to the simple approximation (\ref{QCNLS_solit}) we also observe that at fixed $b$ and $\sigma_r$ it suggests the existence of the upper limit,  $\sigma_i\leq \si^{upp}$, of the strength of the dissipative term  $\sigma_i$, for which localized dissipative solitons exist. This is indeed confirmed numerically in Fig.~\ref{fig3}(a) [notice that the  simple estimate for this upper limit $\si^{(upp)}\approx 3\sr^2+3/4$ gives for the parameters of Fig.~\ref{fig3}(a) $\si^{upp}\approx 1.22$ while the numerical value is $\si^{upp}\approx 1.62$]. The growth of $\si$ results in the monotonic increase of the imaginary part of the field [c.f. also Figs.~\ref{fig1}(a) and (b)]  accompanied by a considerable increase of current density [Fig.~\ref{fig1}(d)]. The energy flow increases with $\si$ [Fig.~\ref{fig3}(a), red curve] until the tangential line to $U(\si)$ becomes vertical.
Apparently, there exists another upper branch of solutions joining with the lower branch in the point $\si=\si^{upp}$  [Fig.~\ref{fig3}(a), black curve] for which the energy flow is a monotonically decreasing function of $\si$. The solitons belonging to this branch are characterized by a double-hump field modulus profile [Fig.~\ref{fig1}(c)]. When  $\si$ decreases the real part of the solutions decays and only imaginary survives. The later is  asymmetric and its maximum and minimum are located in a single period of $V(\eta)$ [this tendency is visible in Fig.~\ref{fig2}(c)]. The solitons from upper branch in Fig. ~\ref{fig3} (a) are unstable. Besides these simplest branches one can find a variety of soliton families with more complicated internal phase distributions, but we do not discuss them here because they are usually unstable.

One of the most important results of this Letter is that fundamental solitons
can be stable despite the fact that the system (\ref{CNLS}) is characterized by the presence of domains where only losses or gain are acting. The outcome of stability analysis is presented in
Figs.~\ref{fig3} (b)-(d). The fundamental solitons are stable for $\si$ below certain critical value $\si^{cr}$ [see Fig. ~\ref{fig3} (d) for a typical dependence of the  perturbation growth rate on $\si$]. Notice that for $\si>\si^{cr}$  the growth rate $\delta_r$ increases until one reaches the border of existence domain $\si=\si^{upp}$. For fixed $\sr$ the stability domain on the plane $(b,\si)$ is rather complex [Fig. ~\ref{fig3} (c)]. At $\sr=0.5$ both $\si^{cr}$ and $\si^{upp}$ increase as $b\to0 $ indicating on soliton stability in a very broad range of amplitudes of gain modulation. For sufficiently large $b$  values the domains of existence and stability monotonically expand with $b$. A similar situation is encountered for other values of  $\sr$.
The increase of the depth of modulation of conservative nonlinearity $\sr$ at fixed $b$ results in considerable expansion of existence domain on the
plane $(\sr,\si)$.

However, especially interesting situation occurs at $\sr=0$.  In this case, there is no modulation of conservative nonlinearity at all, but our analysis still predicts stability of fundamental solitons between two red lines in Fig.~\ref{fig3}(c) (for $b>b^{cr}\approx 1.05$ the solitons are stable for $0<\si<\si^{cr}$). This fact is really remarkable taking into account that now the symmetric conservative nonlinearity    providing the restoring force in the case of slight displacements of soliton center from the equilibrium position $\eta=0$, is absent.
We observe that the loss of stability occurs at the soliton width, which is comparable with the characteristic scale of the lattice, i.e. to the half-period $\pi/2$. 
Thus, the modulation of conservative nonlinearity is not a necessary ingredient for soliton stability, although it can change considerably stability properties of low-power solitons with $b\to0$.

\begin{figure}
\vspace{-1.7cm}
\includegraphics[width=\columnwidth]{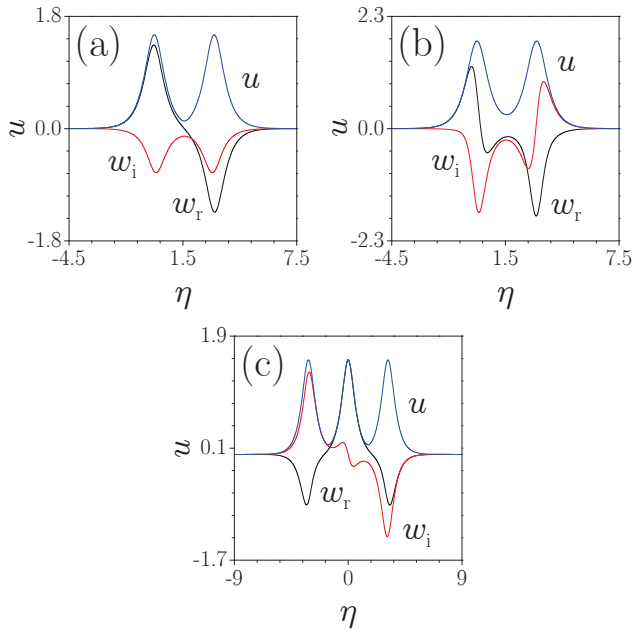}
\vspace{-1.5cm}
\caption{ (Color online)  Dipole solitons at (a) $\si=0.5$, (b) $\si=2.67$, and tripole solitons at (c) $\si=1$. In all cases $\sr=1$, $b=2$}
\label{fig4}
\end{figure}

\begin{figure}
\includegraphics[width=\columnwidth]{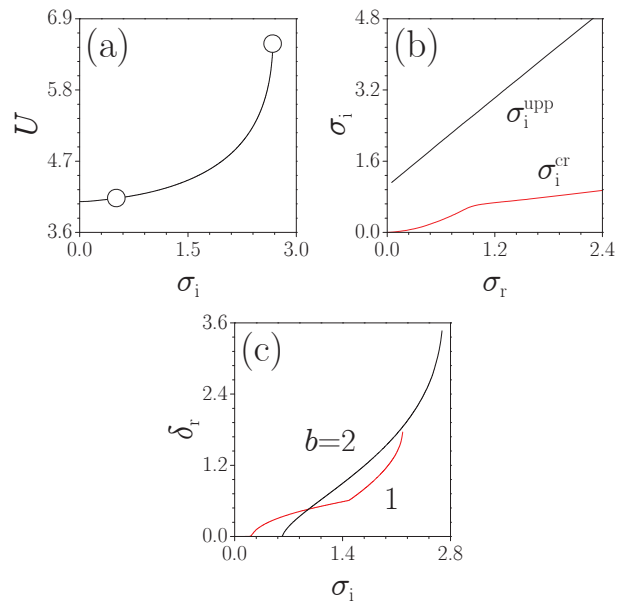}
\vspace{-1.7cm}
\caption{ (Color online)  (a) Energy flow {\it vs} nonlinear gain for dipole solitons at $\sr=1$, $b=2$. Only lower branch is shown. Circles correspond to profiles shown in Figs.~\ref{fig4}(a) and  (b). (b) Domains of existence and stability on the plane $(\sr,\si)$ for dipole solitons at $b=2$. (c) Real part of perturbation growth rate {\it vs} the gain parameter at $\sr$.}
\label{fig5}
\end{figure}

In addition to the fundamental solitons we found a variety of multi-hump states whose humps reside on different maxima of conservative nonlinear lattice $V$. The representative examples of such states that in the limit $\si\to0$ transform into conventional dipole and tripole solitons are shown in Fig.~\ref{fig4}. In such solitons the real part of the field (dominating at $\si\to0$) changes its sign between neighboring maxima of $V$. The current density in such states is characterized by $n$ ($n$ is the number of peaks in field modulus) negative spikes in the vicinity of maxima of $V$. Analogs of solitons with in-phase field peaks were obtained too, but they all are unstable. Like fundamental solitons, multipole states are parameterized by the propagation constant $b$. For a given $\sr$  there exist a cutoff on $b$ below which multipole solitons do not exist, while increase of $b$ results in growth of energy flow.

Increase of gain-loss modulation $\si$ also causes increase of $U$ and fraction of power concentrated in imaginary part of the of multipole soliton [c.f. Figs. ~\ref{fig4}(a) and (b)], but such solitons can be found only at $\si<\si^{upp}$ [in Fig.~\ref{fig5}(a) we show only the lower branch of dipole solitons although upper unstable branch can be found too]. Linear stability analysis predicts stability of the multipole solitons at $\si<\si^{cr}$ as shown in Fig.~\ref{fig5}(c). This domain of stability gradually broadens with increase of the depth of modulation of conservative nonlinearity $\sr$ [Fig.~\ref{fig5} (b)]. In contrast to fundamental solitons multipole solitons can be stable only if propagation constant is sufficiently large. This critical value of propagation constant increases with decrease of $\sr$. This is because multipole solitons may exist only if conservative nonlinearity is modulated and when this modulation is sufficient for compensation of repulsive forces acting between neighboring poles. Increase of the number of poles in solitons does not result in dramatic modifications of existence domain $[0,\si^{upp}]$  but domain of stability $[0,\si^{cr}]$ shrinks with $n$.

To conclude, we have reported a set of stable localized solutions supported by \PT-symmetric nonlinear lattices. The system considered reveals a number of unusual properties. First, although it is dissipative  and the balance between the gain and losses must be satisfied, it possesses families (branches) of solutions, which can be parametrized by the propagation constant $b$, in contrast to other typical dissipative systems. Second, the modes, whose width is smaller  than the lattice  half-period appear to be remarkably stable, even when the conservative nonlinear  potential is absent. Finally, the system supports stable multipole solutions.

The work of FKA and VVK was  supported by
the grant PIIF-GA-2009-236099 (NOMATOS). DAZ was supprted
by the grant SFRH/BPD/64835/2009.

\end{document}